# Generalized Virtual Networking: an enabler for Service Centric Networking and Network Function Virtualization

Stefano Salsano[(1)], Nicola Blefari-Melazzi[(1)], Francesco Lo Presti[(1)], Giuseppe Siracusano[(1)], Pier Luigi Ventre[(2)]

(1) University of Rome Tor Vergata - (2) Consortium GARR

*Abstract*— In this paper we introduce the Generalized Virtual Networking (GVN) concept. GVN provides a framework to influence the routing of packets based on service level information that is carried in the packets. It is based on a protocol header inserted between the Network and Transport layers, therefore it can be seen as a layer 3.5 solution. Technically, GVN is proposed as a new transport layer protocol in the TCP/IP protocol suite. An IP router that is not GVN capable will simply process the IP destination address as usual. Similar concepts have been proposed in other works, and referred to as *Service Oriented Networking*, *Service Centric Networking*, *Application Delivery Networking*, but they are now generalized in the proposed GVN framework. In this respect, the GVN header is a generic container that can be adapted to serve the needs of arbitrary service level routing solutions. The GVN header can be managed by GVN capable end-hosts and applications or can be pushed/popped at the edge of a GVN capable network (like a VLAN tag). In this position paper, we show that Generalized Virtual Networking is a powerful enabler for SCN (Service Centric Networking) and NFV (Network Function Virtualization) and how it couples with the SDN (Software Defined Networking) paradigm.

I. INTRODUCTION

In order to meet the requirements coming from end-user mobility and from service mobility in a more and more pervasive cloud computing environment, the concepts of Service Centric Networking or Service Oriented Networking have recently emerged [1][2][3] (in the rest of the paper we choose to use the expression SCN - Service Centric Networking with the widest possible meaning, including all the variants that have been proposed). The common intuition behind the different proposals is that services should be accessed independently from the IP network address (and transport port) of the node that can run the service instance at a given time. Therefore the routing of packets towards a service instance should not be based on IP network address and transport port of the service, but on some higher level information. This need has been exacerbated with the evolution of cloud infrastructures towards "edge computing" or "fog computing" [4], which break the traditional tight separation between computing and networking resources by dispersing computational capacity inside the network.

The recently proposed Network Function Virtualization paradigm [5][6] foresees to implement networking functions as a chain of NFV processing elements that can be executed on a distributed processing environment. We observe here that the problem of routing a packet through the chain of NFV processing elements has a strong commonality with the routing of a packet toward a service instance in Service Centric Networking. Even further, we can see NFV as a special case of Service Centric Networking.

The SCN proposals in [1][2][3] agree in the approach of leaving the IP layer unchanged and operate above the networking layer. Considering the current trends in IP networks, we agree with this approach. In particular, we believe that it would be useful to generalize the different proposal with a common supporting framework that can help turning these proposals into reality. Moreover, we believe that the same framework can also serve the needs of Network Function Virtualization. Hence in this position paper we introduce the concept of GVN - Generalized Virtual Networking. GVN allows to add information to IP packets that can be processed by GVN aware end nodes and network nodes to influence the routing of the packets. It does not prescribe *how* to route the packets and *what* type of SCN information should be carried, but it provides a structured approach for the coexistence of different SCN and NFV proposals. The implementation and deployment of GVN can greatly improve the efficiency of SCN processing both in end-nodes and network nodes, can ease the interaction with Software Defined Networking (SDN) architectures and can favor the introduction of innovative SCN solutions in IP networks.

We started our work on GVN with the consideration that that the vast majority of today's networked services and applications are realized using an overlay approach, overcoming the inherent limitations of the Internet protocol and architecture. Examples include Peer-to-Peer file sharing systems, Content Delivery Networks, integrated platforms for VoIP/video/messaging like Skype and many others. All these services and applications sit on top of IP protocol as "overlays" and mostly operate at the application level. The networking level (IP) is not modified even if this could bring improvements to the offered service level or to the efficiency of service provisioning. Several attempts have been carried out trying to counteract this trend, extending or redesigning the networking layer so as to include better built in support for advanced features like name-based or content-based communications, terminal mobility, service mobility. At the time of writing all these attempts have failed to break the boundaries of laboratory experimentation and start being deployed in the wild. On the other hand, cross-layer networking is a reality and a plethora of "middlebox" devices interacts with IP packets at levels above the networking layer. Services and features like Network Address Translation, Firewalls, Load Balancing, Deep Packet Inspection and others are implemented by middleboxes that inspect and sometime modify packet headers potentially at all levels of the protocol stack. This has been done on a case-by-case approach leading to a huge complexity and high cost of development and maintenance of this type of solutions. As we will show, GVN "institutionalizes" overlay networking and cross-layer networking which are already standard practices in IP networking, making easier for end nodes and network nodes to process and mangle the packet headers.

The recently proposed Software Defined Networking (SDN) concept has introduced a separation of control and data plane, allowing a *SDN controller* to inject forwarding rules in "dumb" switching equipment. The SDN forwarding rules in switches are based on the concepts of *matches* and *actions*. A match specifies a set of conditions related to the content of the packet headers, an action specifies if and how to modify the content of the packet headers and where to forward the packet. This is typically implemented in a cross- layer fashion: in principle, the forwarding rules can concern packet headers at all levels of the stack from layer 2 to application layer. The current specification of the OpenFlow protocol [7], the most prominent implementation of the SDN concept, allows to match and operate on header fields from layer 2 up to transport layers (e.g. TCP or UDP ports). The separation of control and data plane brought in by the SDN concept can be of great benefit for the proposed or future SCN architectures. Therefore it is important to consider how SDN can be applied to SCN so as to foster and ease the innovation. In our opinion, the most critical part is the support of arbitrary headers and packet formats that can be introduced by SCN architectures especially if operating as overlays. It will be difficult to extend the match/action capabilities of the switches and to reflect these extensions on the switch-to-controller protocol (e.g. OpenFlow) following the arbitrary needs of different SCN architectures. Having a framework like the proposed GVN that allows to carry SCN related headers in a structured way does not automatically solve the issue, but it is of great help to define a future-proof SDN solution capable to serve the needs of Service Centric Networking.

The proposed Generalize Virtual Networking is based on a protocol header between the network (IP) and transport headers. Hence, GVN is a transport protocol carried in IP packets. An IP router node that is not GVN capable will simply process the IP destination address as usual. A node that is GVN capable will inspect the GVN header and will be able to process the packet considering the information contained therein. In this respect, GVN can be considered a *level 3.5* protocol, as for example MPLS is considered a *level 2.5* protocol. In [1] a similar approach has been proposed and called Service Centric Networking, while in [3] it has been called Application Delivery Networking. The idea of GVN is to generalize the set of existing proposals in a common framework which can be particularized to the different specific approaches and use cases. We call these different approaches as GVN Processing Logics (GVN-PLs). The GVN framework allows different innovative GVN-PLs to smoothly coexist in the protocol stacks of network devices and hosts.

We believe that such a common framework can foster innovation and facilitate the introduction of SCN concepts and features in real networks. The advantages of GVN will materialize if the GVN framework will be implemented and available in end-nodes and network node, offering APIs with library and software tools for handling SCN information. This drastically simplifies the introduction of innovative arbitrary Service Centric Networking solutions. As a follow up of this position paper, we are starting the development of an open reference implementation of GVN for the Linux networking stack. Moreover the standardization of GVN will be important to gain widespread acceptance and penetration, though this is not a blocking requirement. In fact, the GVN solution can be *incrementally* deployed and is fully compatible with legacy networking devices and end nodes. Waiting for its standardization, GVN can be implemented using the "experimental" codes that been reserved to such purposes.

In section II we review the different proposals that in our opinion can be supported by the GVN framework. In section III we provide the details of the proposed solution. In section V we provide a tentative mapping of some proposed solutions (described in section II) over the GVN framework, showing how it can become an universal enabler for Service Oriented Networking. In section VI we discuss the relation between GVN, NFV (Network Function Virtualization) and SDN (Software Defined Networking), showing how GVN is a powerful enabler for NFV and couples very well with the SDN paradigm.

## II. REQUIREMENTS AND STATE OF THE ART

In this section we analyze a set of proposed Service-Centric / Information-Centric Networking architectures from which we have derived the requirements for the GVN solution. In section V we give some further hints at how GVN can support the considered architectures.

### A. Service Centric Networking

#### 1) Fusion

The concept of Service Oriented Networking [2] has been developed in the context of the EU project FUSION. The idea is that networked software functions are dynamically deployed, replicated and invoked, similarly to what is proposed for static content in Information Centric Networking [8][9]. The mechanisms required to deploy a replicated service instance in the network and to route client requests to the closest instance in an efficient manner are discussed in [2]. At the networking level, the need for a service-anycast capability is advocated, capable to optimize service instance selection based on network metrics as well as server load. In Service Oriented Networking proposed by FUSION, services are identified by a *serviceID*. For the anycast routing of requests towards service instance, the options of a clean slate approach and of extending the DNS were analyzed and discarded. An overlay routing solution has been selected, with the main advantage of being easily deployable. It offers two options: i) ServiceIDs can be resolved into IP addresses by the overlay routing layer, allowing direct IP level communications between invoking host and the service instances, ii) the overlay routing layer can directly forward application data (e.g. for simple request/response transactions).

*2) Serval*

The Serval architecture [1] has been designed in order to meet the requirements of current online services, running on multiple servers in different locations and serving clients that are often mobile and multi-homed. There is a clear mismatch between these requirements and today's network stack, designed for communication between fixed hosts with topology-dependent addresses. Starting from this observation, Serval proposes a Service Access Layer (SAL) that sits above an unmodified network layer (IP), and enables applications to communicate directly on service names, instead of using IP addresses and transport level ports. Unlike traditional "service layers," which sit above the transport layer, the SAL is located between the network and the transport layer. It offers a programmable service-level data plane that can adopt diverse service discovery techniques. Serval proposes a serviceID, that corresponds to a group of one or more processes offering the same service. The Serval protocol header, introduced between the IP and the transport layer headers, is shown in [1]. It consists in two groups of fields, named "Service access" and "Service access extensions" (which includes the serviceID). The socket abstraction between applications and transport/ network layers is redefined by Serval using the serviceID as end-point identifier instead of the IP address and port number used by the current socket interfaces. A prototype of Serval, implementing the Service Access Layer has been realized. The SAL can operate on end-hosts only or on intermediate nodes capable of Serval forwarding called "service routers" (SR).

*3) OpenADN*

The definition of the OpenADN architecture starts from the consideration that the *service-centric* delivery semantics of modern Internet-scale applications and services does not fit naturally into the Internet's host-centric design. To address this gap, Application Service Providers deploy a service-centric logical infrastructure using mechanisms like data-plane proxying and Layer 7 traffic steering. This is expensive and difficult to achieve in wide area distributed data centers. OpenADN aims at providing a general architectural support for service-centric Internet. It is worth noting that the OpenADN design leverages the Software Defined Networking paradigm to implement and manage the OpenADN services. According to [3], OpenADN provides an application-neutral, standardized, session-layer overlay over IP. It introduces two layers in the protocol stack, one between the network and the transport layer and another one on top of the transport layer, offering the API towards the applications. Accordingly, as shown in [3], a layer 3.5 header in introduced between the IP header and the transport headers and a layer 4.5 header is introduced after the transport headers. The OpenADN data plane implements an MPLS inspired label switching and stacking mechanism called APLS (APplication Label Switching). Hence the OpenADN protocol headers are denoted as *L3.5 APLS Header* and *L4.5 APLS Header*. The OpenADN architecture considers end-user hosts, middleboxes and application servers that can exchange message through *OpenADN Proxies*. A prototype of OpenADN has been realized based on the click modular router [12].

*B. Information Centric Networking*

The Information Centric Networking (ICN) concept [9] is proposed as a paradigm shift from the host-to-host communication model of current TCP/IP networks to a model that focuses on information objects and their distribution. Also known as CCN (Content Centric Networking) [8], it has been initially proposed among the "clean-slate" approaches aiming at redesigning network architectures getting rid of the TCP/IP legacy. We observe that several of the implemented ICN inspired architectures use an overlay approach over IP or try to extend TCP/IP network architecture rather than replacing it from scratch, see for example [14] [15]. In doing so, these ICN proposals need to carry their protocol headers in one of these ways: i) within UDP or TCP transport protocols (overlay approach); ii) by defining new transport layer protocols; iii) by extending the IP layer (e.g. with new IP header options). We advocate that a generic and extendible framework to support the exchange of ICN control messages and data in the TCP/IP suite could be beneficial to ease the development and the deployment of ICN based solutions.

We have worked on a generalization of ICN architectures called Internames [13]. In the Internames architectural framework names are used to identify all entities involved in communication: contents, users, devices, logical as well as physical points involved in the communication, and services. Internames enlarges the scope of ICN, extending its functionality beyond content retrieval, easing send-to-name and push services, and allowing the use of names to route data also in the return path. In this respect, Internames can be seen both as an Information Centric Networking architecture and as an enabler for Service Centric Networking.

Finally we would like to mention our experience with the implementation of ICN over SDN [16] as a source of requirements for GVN. In [16] we have considered the ICN architecture called CONET [14][15] which extends the CCN proposal [8] and investigated how to support it with a SDN based approach. We designed and implemented a solution suited to operate with the OpenFlow implementations available in switches and controller as of 2013 (i.e. OpenFlow 1.0, one of the previous versions of [7]). We suffered from the rigidity of header matching capabilities offered by the switches and supported by the OpenFlow protocol. Therefore our solution is based on (mis-)using the UDP header by carrying a content identifier tag in place of UDP source and destination ports. The tag has to be inserted and then removed at the edges of an "ICN over SDN" capable domain. We realized the importance of specifying a generic mechanism to insert Information Centric (or Service Centric) related headers, so that in a medium term perspective the switch can implement and offer adequate capabilities for matching /inspecting / modifying the content of these headers.

## III. GVN - Generalized Virtual Networking

Before introducing the Generalized Virtual Networking solution, we would like to discuss some features of VLAN and MPLS, two of the most successful paradigms in the networking arena. In fact, VLAN provides the foundation of current networking in large scale layer 2 Local Area Networks, now often extended to a geographical scope in Metropolitan Area Networks and in distributed data centers. MPLS is the solution of choice for the large majority of IP network providers for their IP backbones.

Both VLAN and MPLS share the approach that an additional portion of the packet header can be inserted in a pre-existing packet when needed and then removed when it is not needed anymore. This is commonly referred to as "tagging" and "un-tagging" (e.g. VLAN tagging). This preserves compatibility with legacy devices that are not VLAN or MPLS aware *outside* of a VLAN or MPLS island. In this respect, VLAN does something more, as it enables legacy layer 2 devices to process a VLAN tagged packet as a regular Ethernet packet, without considering the VLAN header. Therefore, compatibility with legacy devices *inside* the VLAN island is even ensured. In principle the VLAN and MPLS tagging and un-tagging can be performed by end hosts or by intermediate devices. In the VLAN case, both options are used in the current networking practice, while in the MPLS case only the option of tagging/un-tagging by intermediate devices (called MPLS Edge Routers) is commonly used. The VLAN technology is mainly used to enforce a virtual separation of networks insisting on the same layer 2 network, MPLS can used for the same purpose with the different types of MPLS VPNs (Virtual Private Networks) but it operates in layer 3 networks. MPLS is also used with the purpose of explicitly influencing the routing of packets, something which is not usually done with VLANs.

Generalized Virtual Networking combines most of the above discussed features of the successful VLAN and MPLS technologies. The position of GVN header in the packet header is shown in Fig. 1, with reference to IPv4. The GVN header can be inserted between the IP and transport header and then removed when it is not needed.

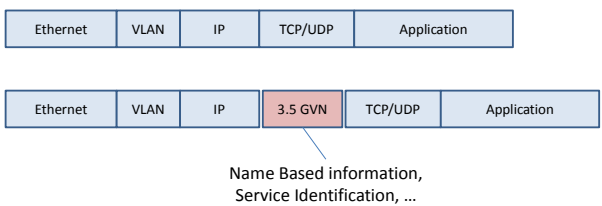

Fig. 1. "level 3.5" GVN protocol header for IPv4

Depending on the scenario, the GVN header can be added/removed directly by a GVN-capable end-hosts or by an intermediate devices. Similarly to the VLAN approach, GVN tagged packet can be processed (at IP level) by intermediate nodes (i.e. IP routers) that are not GVN capable. The GVN header can be used: i) to provide Virtual Private Networks enforcing virtual separation of networks insisting on the same IP network; ii) to explicitly influence the routing of packets, like in the MPLS approach.

A significant difference between VLAN and MPLS tags and GVN header is that the GVN header can have a variable length. A second important difference is that the semantic of the GVN header is not known in advance (apart from the first 8 bytes). The format of the GVN header is shown in Fig. 2. The GVN header starts with a 1 byte long field called GVN length. It represents the length of GVN header in multiple of 4 bytes. The minimum value of GVN length field is 2 as the minimum length or GVN header is 8 bytes. The GVN header length can range from 8 to 1016 bytes in steps of 4 bytes, (as the value of 255 for the GVN length field is reserved). The second header byte carries the Next header The third header byte carries a "Flags" field reserved for future specification. Then there is a 5 bytes field called "GVN code". The GVN code defines how to interpret the rest of the GVN header, providing the association with a GVN Processing Logic (GVN-PL). It allows different semantic interpretations of the GVN header, as well as the coexistence of different independent GVN-PLs exploiting the GVN framework (e.g. different Service Oriented Networking proposals, different Information Centric Networking proposals, …). The GVN header transports a variable length "PL-Specific Header Data" field whose semantic is defined by the GVN Code field. GVN Codes (and code blocks) can be allocated to organizations like MAC addresses or assigned to a given usage by Standard Defining Organizations. A coordinating organization is needed in any case to regulate the assignment of GVN codes, like IEEE does for Ethernet "MAC" addresses. This aspect is shortly discussed further in section IV.

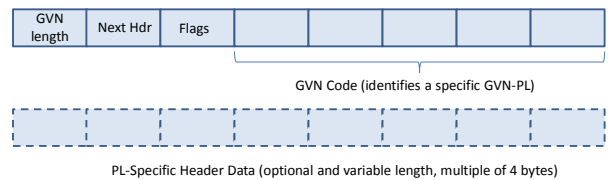

Fig. 2. GVN header format

In case of IPv6 the GVN protocol extension is even more straightforward and clean. IPv6 header has been designed to allow the introduction of new "Extension Headers" that are meant to enhance the networking layer. In fact the difference between a network layer extension and a transport layer header is somehow blurred in IPv6. The GVN extension header will be introduced before the transport layer header and its Next Header field will refer to the transport header included in the packet.

In the TCP/IP protocol suite, the IP protocol type code carried in the IP header is used to distinguish the different transport protocols (e.g. 6 for TCP, 17 for UDP). GVN is seen as a transport protocol, therefore a new IP protocol type will need to be assigned by IETF to GVN to complete its standardization. Note that IP protocol type codes are shared between IPv4 and IPv6, therefore only one code needs to be assigned and it will serve for both versions. Waiting for the standardization, one of the two IP protocol type codes reserved for experimentation can be used for prototype implementation (253-254 or 0xFD-0xFE in hexadecimal), we choose to use 254/0xFE).

## A. Processing of incoming GVN packets

Let us now describe the processing of GVN tagged packets by receiving IP nodes. We considering the four possible combinations of these two aspects: i) the receiving node can be a legacy node (*LEGACY*) or a GVN capable node (*GVN*); ii) the IP destination address of the packet corresponds to a IP address of the receiving node (*IPnode*) or not (*IPother*).

(*LEGACY*) If the receiving IP node is a legacy node, it is obviously not able to understand the GVN header, as it will have no notion of the GVN transport protocol. In this case it will simply process the packet at IP level using the IP destination address. If the destination IP address does not correspond to a local IP address of a node interface (*LEGACY.IPother*), the IP node will forward it according to its IP routing table. This will be the behavior of a legacy IP router processing of a GVN packet. If the destination IP address is a local IP address of the node (*LEGACY.IPnode*), the node will try to process the transport protocol and, in this hypothesis, it will find an unknown transport protocol and will have to discard the packet (in general, this will be the result of an error condition).

(*GVN*) If the receiving node is GVN capable it will be able to process the first 8 bytes of the GVN header. In particular, it will read the 6 bytes GVN code. This code univocally identifies how to interpret the rest of the GVN header. If the receiving node is not able to understand a GVN code it will just process the packet at IP level using IP destination address. The processing will be exactly as described above for the (*LEGACY*) case. If the receiving node is able to understand the GVN code, it means that it can process the packet according to a specific semantic and service logic. Obviously the packet processing will take into account the IP destination address of the packet if needed, covering the (*GVN.IPnode*) and (*GVN.IPother*) cases as appropriate. The flag field in the GVN header could be used to standardize some common processing behaviors. We are still in the process of analyzing the possible requirement, as an example one bit of the flag field could indicate whether a GVN capable node should forward or not a packet if an unknown GVN code is found.

## B. The GVN code as the key to open innovation

As outlined above, apart from the first 8 bytes, the semantic of the GVN header is left to the different GVN Processing Logics (GVN-PLs). Completely independent GVN-PLs can coexist in the network thanks to the GVN codes, which provides the mapping to the Processing Logic to be executed on a given packet. A GVN capable node can be seen as an upgradable collection of processing and forwarding behaviors (GVN-PLs) that can be used to support different Service Oriented Networking solutions (as well as Information Centric Networking solutions) as long as the common format of the GVN header is used by all these solutions.

## C. End Nodes GVN scenario vs Edge Nodes GVN scenario

Without the intent to put any limitation on different usages of the GVN framework, which is open to any further innovation we describe two example GVN scenarios, represented in Fig. 3. In the first scenario, referred to as "End hosts GVN" scenario, the GVN header is managed by GVN capable end-hosts and applications. In the second scenario the GVN header is pushed/popped at the edge of a "GVN enabled domain" by GVN Edge Noes (like the MPLS tag is pushed/popped by MPLS edge nodes at the border of a MPLS domain). The two scenarios are complementary and they can coexist in different GVN PLs (or even in the same PL).

## D. An open issue: do we need NAT traversal?

A potential limitation of the proposed GVN transport level header is that the NAT traversal of a packet with a GVN transport header through a legacy NAT will not be supported. We need to investigate further if there are scenarios in which this fact can represent a limitation, and which countermeasures can be taken. For example, if the GVN enabled nodes are placed after the legacy NAT boxes there is no problem in communicating with the end-nodes which sit behind the NAT, as the GVN header will not be inserted in the packets that cross the legacy NAT. Likewise, if the NAT service is provided by a GVN enabled solution, there is no problem because there will be no legacy NAT involved. The only critical case is when a set of GVN enabled nodes in a private network environment wants to communicate with other GVN enabled nodes in the Internet or in other private network environments, keeping a transparent connection from the point of view of the GVN Processing Logic operating on top of GVN. If the support of this scenario will be considered a useful requirement, solutions can be designed, based for example on tunneling solutions capable of NAT traversals.

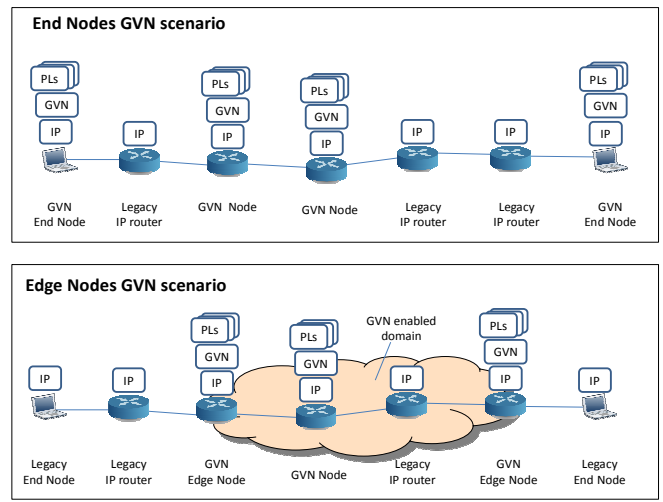

Fig. 3. Example GVN scenarios (End-Nodes GVN vs Edge Nodes GVN)

## E. Generalized Virtual Networking at layer 2

In this subsection we shortly mention that the same approach that we have used for GVN as a transport protocol over IP could be done at layer 2, realizing a 2.5 GVN approach. In this case an Ethertype would need to be assigned to GVN, which will become a layer 3 protocol over Ethernet. The Ethernet payload would start with the GVN header, which could be defined with the same format described in Fig. 2 (more specifically, a two byte field for the Next Header information would be needed as Ethertypes are two bytes long). The same processing rules described in section III.A can be applied, by replacing the IP destination address with the MAC destination address. A node that is not layer-2-GVN capable will process the packet using the MAC destination address. A node that is layer-2-GVN capable will check the GVN code to associate a GVN-PL. If the GVN code does not correspond to a known Processing Logic the packet will be processed using the MAC destination address.

Deploying GVN at level 2.5 could be an alternative to the proposed 3.5 GVN approach, with the advantage that also other protocols than IP could be supported and that there will be no need of separate design and implementations for IPv4 and IPv6. On the other hand we currently see no strong needs for the support of other layer 3 protocols and we judged that implementing GVN at layer 3 is easier than doing it at level 2. For these reasons, we are not further pursuing this layer 2.5 approach for implementation and standardization and just focusing on 3.5 GVN.

## IV. STANDARDIZATION ASPECTS

The standardization of GVN could be performed in the context of IETF. An IP protocol type code needs to be assigned, and the format of the GVN header, as for example proposed in Fig. 2, should be agreed. Once the framework has been standardized, IANA could take care of supervising the assignment of GVN codes to organizations.

## V. GVN AS UNIVERSAL ENABLER

We claim that GVN can support an ordered coexistence of several Service Centric or Information Centric Networking architectures. GVN can act as a development facilitator and as a deployment framework, to have the different proposal sent out to the wild and let the strong survive by natural selection. The advantage of such coordinated approach is evident if we only consider the two exemplary SCN proposals (Serval and OpenADN) described in section II.A. Both propose a new protocol header between IP and transport layer, which in fact requires the standardization of a new IP protocol type. This would imply to have a different IP protocol type code for each different SCN proposed architecture, while the numbering space is relatively small (256 codes, of which 146 have been allocated and 110 are still available). With the GVN solution, a single IP protocol type code would serve the need of a virtually unlimited number of SCN solutions (with $2^{40}$ available GVN codes).

Based on the information and the figures available in [1] and [3], in Fig. 4 we have tried to illustrate how Serval and OpenADN headers could respectively be mapped into the proposed GVN protocol header. As for the Serval mapping, the Transport Protocol field is redundant if it indicates the internal transport protocol (already carried in the NextHdr field of the GVN header). Therefore we have inserted it in gray italic font in the figure. As for the OpenADN mapping, we considered the APLS 3.5 header, which surely is of interest of both end-nodes and network nodes. It has to be considered if it could be useful also for the APLS 4.5 header to be moved from the application layer and transported inside the GVN header.

Coming to the support of ICN architecture, just as a simple example we depict the mapping on GVN of our solution for ICN over SDN [16] in Fig. 4. Without entering in the details, the PL-Specific Header Data field includes a fixed length tag that uniquely represents a content object. We note here that the PL-Specific Header Data field can be structured in a suitable way to ease the design and implementation of proper SDN match/action mechanisms.

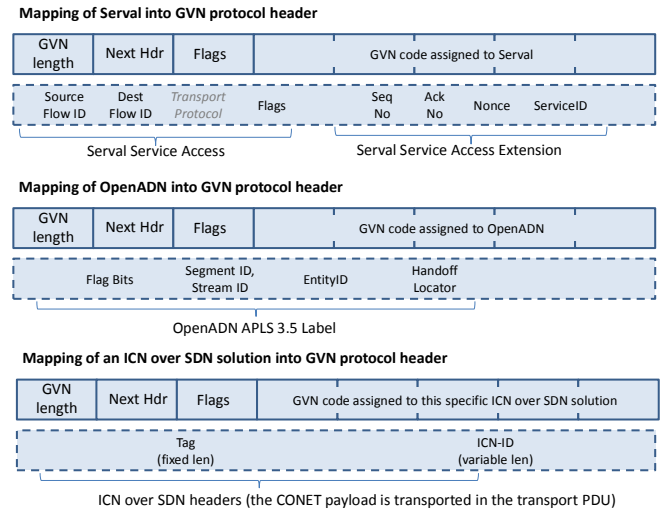

Fig. 4. Mapping different SCN /ICN solutions on GVN

## VI. GVN, NFV AND SDN

The Network Function Virtualization concept [5][6] proposes to virtualize the network node functions in building blocks that can be executed in distributed environments (e.g. data centers) and that can be chained to realize complex network services. The IETF is now addressing the NFV concept within the Service Function Chaining WG, which has produced a problem statement document [17].

In a NFV enabled architecture, the routing of IP packets across a chain of potentially distributed NFV elements is not directly related to the IP destination address and transport port of the packet, but it should be controlled by the NFV service logic. Therefore this problem can be seen as a particular case of Service Centric Networking and our position here is that it can be successfully supported by the proposed Generalized Virtual Networking.

A solution to NFV routing issue has been proposed, called Network Service Header (NSH) [18]. In the NSH solution, a service header called NSH is added to a packet (typically an IP packet), then the NSH and the original packet are encapsulated in an outer packet (e.g. using GRE or VXLAN tunneling). The NSH header is meant to contain metadata and service path information needed by the NFV routing, in a way that can be easily extended and adapted to the service requirements. We advocate that in case of IPv4 and IPv6 packets, a solution based on GVN can be designed and used to carry NSH information, with no need to encapsulate the packets. In particular, in order to have the same functionality of the NSH solution based on encapsulation, the IP destination address may have to be rewritten and the original destination address saved in the GVN header. In this way the packet can be routed to any NFV enabled node using legacy IP intermediate nodes, and the original destination address can be restored at the exit of the NFV chain.

As for SDN, we remark the relation with GVN. The idea is that SDN can be used to control a GVN enabled network, using forwarding rules based on the GVN header in GVN capable and SDN enabled nodes. This does not come for free, as the current SDN enabled nodes are clearly not capable of GVN, nor the switch-to-controller protocols (e.g. OpenFlow) support GVN. Our position here is that the deployment of the GVN framework can ease the integration of Service Centric Networking with SDN, with respect to the current situation where several independent solutions are proposed for SCN, arbitrarily introducing SCN information at all the levels of the protocol stack, from application level (overlays) to transport level, to IP and even layer 2.

## VII. IMPLEMENTATION ASPECTS

As already pointed our discussed in [3], a layer 3.5 solution like GVN is expected to be implemented as a kernel module in the end-nodes and network nodes Operating Systems for performance reasons. This is clearly more challenging than the implementation of a Service Centric Networking solution using an overlay approach. What we plan to do in our implementation is to develop a kernel module for the basic GVN support in the Linux OS. This module can work for a Linux based end-node as well as for a Linux based network node. The GVN Linux kernel mode will offer the chance to plug-in a specific GVN Processing-Logic both as a kernel module and as a user space plugin.

## VIII. CONCLUSIONS

In this position paper, we have presented the Generalized Virtual Networking (GVN) solution and shown how it can fully support the needs of Service Centric Networking and of Network Function Virtualization. GVN is meant as a framework to support different specific solutions, facilitating the introduction of innovative architectures in IP networks. In the coming time, we plan to provide a reference implementation of GVN and to propose it to relevant standardization fora like IETF.

## ACKNOWLEDGMENTS

This work has been partly funded by the EC in the context of the "DREAMER" project, one of the beneficiary projects of the GÉANT Open Call research initiative, and of the "GreenICN" joint EU-Japan project.